\documentclass[conference]{IEEEtran}
\IEEEoverridecommandlockouts
\usepackage{cite}
\usepackage{amsmath,amssymb,amsfonts}
\usepackage{algorithmic}
\usepackage{graphicx}
\usepackage{textcomp}
\usepackage{xcolor}
\usepackage[utf8]{inputenc}
\def\BibTeX{{\rm B\kern-.05em{\sc i\kern-.025em b}\kern-.08em
    T\kern-.1667em\lower.7ex\hbox{E}\kern-.125emX}}

\newcommand{\UTFPR}{\small{Federal University of Technology - Paraná}}

\begin{document}

\title{Short Datathon for the Interdisciplinary Development
	of Data Analysis and Visualization Skills\\
}

\author{\IEEEauthorblockN{Myrian Noguera Salinas}
\IEEEauthorblockA{\textit{Academic Department of Informatics} \\
\textit{\UTFPR}\\
Curitiba, Brazil \\
michinoguera@gmail.com}
\and
\IEEEauthorblockN{Maria Claudia Figueiredo Pereira Emer}
\IEEEauthorblockA{\textit{Academic Department of Informatics} \\
\textit{\UTFPR}\\
Curitiba, Brazil \\
mcemer@utfpr.edu.br}
\and
\IEEEauthorblockN{Adolfo Gustavo Serra Seca Neto}
\IEEEauthorblockA{\textit{Academic Department of Informatics} \\
\textit{\UTFPR}\\
Curitiba, Brazil\\
adolfo@utfpr.edu.br}
}

\maketitle

\begin{abstract}
Understanding the major fraud problems in the world and interpreting the data available for analysis is a current challenge that requires interdisciplinary knowledge to complement the knowledge of computer professionals. Collaborative events (called Hackathons, Datathons, Codefests, Hack Days, etc.) have become relevant in several fields. Examples of fields which are explored in these events include startup development, open civic innovation, corporate innovation, and social issues. These events have features that favor knowledge exchange to solve challenges. In this paper, we present an event format called {\em Short Datathon}, a Hackathon for the development of exploratory data analysis and visualization skills. Our goal is to evaluate if participating in a Short Datathon can help participants learn basic data analysis and visualization concepts. We evaluated the Short Datathon in two case studies, with a total of 20 participants, carried out at the Federal University of Technology - Paraná. In both case studies we addressed the issue of tax evasion using real world data. We describe, as a result of this work, the qualitative aspects of the case studies and the perception of the participants obtained through questionnaires. Participants stated that the event helped them understand more about data analysis and visualization and that the experience with people from other areas during the event made data analysis more efficient. Further studies are necessary to evolve the format of the event and to evaluate its effectiveness.
\end{abstract}


\section{Introduction}
In Hackathons, small teams produce working software prototypes in a short time period \cite{Komssi2015}. Hackathons are good for collaboration, experimentation, and learning. According to the objective we can group them in some categories such as Innovation, Pedagogy, Community Engagement and Recruitment.
Regarding Innovation, all Hackathons are a way to challenge participants to present an innovative product \cite{Trainer2016,Karlsen2017,Dinter2017}. 
Hackathons in the Pedagogy category are focused on learning \cite{Nandi2016,Karlsen2017,Anslow2016}.
Community Engagement Hackathons involve a community for a few days in an intensive event focused on solving real world problems \cite{Trainer2016, Munro2015}.
Finally, Recruitment Hackathons provide opportunities for participants to seek employment and for companies to seek talented professionals\footnote{http://archive.is/8Bkrd. Last Access: Jan. 8th, 2019.}\cite{Olson37,Flores2018}.

A Datathon is a kind of Hackathon where the challenge is related to data analysis; people come together, for a certain period of time, to work on problems related to a specific dataset \cite{Anslow2016}. The increase in the amount of open data from governments, organizations and agencies, consequently, brought interest of 
professionals to the exploration of this data, and this lead to the organization of such events \cite{Neto2018, Aboab2016, Kohli2017}. However, there is a need to increase the data literacy in students and professionals \cite{Anslow2016} for these events to become more effective.

The goal of this paper is to evaluate if participating in a Short Datathon can help participants learn basic data analysis and visualization concepts. The contributions of this paper are:
\begin{itemize}
	\item A characterization of the design of a Short Datathon for Tax Evasion;
	\item An evaluation of two case studies of a Short Datathon.
\end{itemize}

\section{Short Datathon Design}

We propose a Short Datathon format to achieve our goal through the exploration, preprocessing and visualization of data related to tax evasion provided by a recent paper \cite {Alstadsaeter2017}. One of the authors of this paper, Gabriel Zucman, provided a database of spreadsheets containing tax evasion data used in that paper, which we call {\em Zucman's data} here\footnote{http://gabriel-zucman.eu/leaks/. Last Access: Jan. 8th, 2019.}.

We assume that a 6 hour long event can be successful if we combine the correct tasks and tools. Our goal is to reach a more varied audience, especially people who do not usually participate in events that last 24 hours or more, as many Hackathons do.

\subsection{Structure of activities and content}
The roles in our Short Datathon are three: instructor, judge and participants. The instructor is the main responsible for the event, and the judge helps the instructor to evaluate the presentations of the participants.

The tasks performed during our Short Datathon are divided into 6 steps (Figure \ref{figEstrutura}):

\begin{figure}[h]
	\centering
	\includegraphics[width=\linewidth]{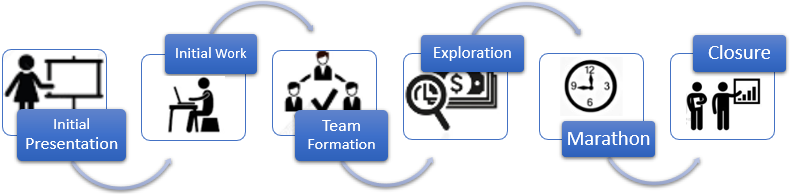}
	\caption{General Structure of a Short Datathon.}
	\label{figEstrutura}
\end{figure}

\begin{itemize}
	\item {\verb|Initial Presentation|}: The instructor presents a contextualization of approximately 45 minutes to deepen topics such as Short Datathon, tax havens, creation of an offshore company, intermediary companies, characteristics of companies and accounts created in tax havens, advantages and disadvantages of offshore companies, a description of the Panama Papers leaks and some cases in Brazil of tax havens usage.  The instructor presents a puzzle activity; by solving it, the participants learn more about the Panama Papers leaks and interact with other participants.
	\item{\verb|Initial Work|}: The instructor presents the functionalities of the data visualization software by using it with Zucman's data. The instructor provides a shared folder\footnote{http://bit.ly/ShortDatathonSharedFolder. Last Access: Jan. 8th, 2019.} with help materials such as: a video showing how to preprocess data, the data visualization software user manual, data already preprocessed to generate graphs and links to suggested data sources.
	\item{\verb|Team Formation|}: Participants split into teams of up to 5 members. The choice of the members for each team is free.
	\item{\verb|Exploration|}: Teams explore data from a variety of data sources.
	\item{\verb|Marathon|}: The teams work with a focus on data preprocessing and graphics generation using Gapminder. 
	\item{\verb|Closure|}: At the end of the event, the teams present the work, the judge and the instructor give verbal feedback, and the instructor closes the event.
\end{itemize}

\subsection{Software for Data Visualization}
A special challenge was to decide which data visualization software was going to be used by the participants. In many Hackathons, the participants choose the tool that best suits them. However, in this work  we decided to define a tool for the following reasons:
\begin {itemize}
\item We needed to prepare the infrastructure of the lab that was going to be used by the Short Datathon. We wanted the software to be already installed on the computers, as it was a short event. Considering the variety of software available for working with data, it would be difficult to install all of them on the computers;
\item We wanted to attract people with little or no knowledge of computer programming. The target audience was students and professionals not only from Computer Science but also from Economics, Business Management, Statistics and others. Therefore, we chose the tool which we believed required less previous knowledge.
\end {itemize}

The software used was Gapminder\footnote{https://www.gapminder.org/. Last Access: Jan. 8th, 2019.}, a tool for visualization of human development statistics created by Hans Rosling. The tool allows you to explore data from various data providers, such as the World Bank, Eurostat and others; with this data, users can create bubble lists, maps, bars and line charts.

\section{Case Study I}

We conducted the first case study to evaluate our Short Datathon
format in the Technological Week\footnote{http://semanatechnologica.dainf.ct.utfpr.edu.br/. Last Access: Jan. 8th, 2019}, an event organized at the Federal University of Technology - Paraná (UTFPR). We used a laboratory with capacity for 25 people, with computers, chairs and a projector. The Short Datathon event was included in the Technological Week
schedule and posted on its website.

The first author, in the role of instructor, started the Short Datathon at 9:20AM with the presence of 12 undergraduate students majoring in Computer Science or Mechanical Engineering. The instructor contextualized the topic, explained the workings of Gapminder and then the participants were divided into teams. The teams carried out an initial exploration of the data and submitted their proposals.

After lunch break, only 6 students returned to continue with the work. This might have happened because in the schedule of the Technological Week, the Short Datathon was published as two separate activities (one in the morning and the other in the afternoon). Then there were people who signed up for other courses or conferences in the afternoon.

At the end of the afternoon the teams presented their work. The "Computer Systems" team was the first to present a chart on HSBC Bank customer figures by country, combined with the
Corruption Perceptions Index (CPI) and the countries fiscal revenue numbers (\% GDP).

The team used three data sources: Transparency Internationals Corruption Perception Index (CPI), Tax Revenue (\% GDP) of Word Bank and Number of HSBC Bank clients in Switzerland from the study carried out by \cite {Alstadsaeter2017}.

In the "X" axis, they plotted Tax Revenue data (\% GDP) and in the "Y" axis the total number of clients with HSBC Bank
accounts by country. They identified the level of Corruption Perception by color intervals; closer to red are countries with high levels of perception of corruption and closer to the opposite green, less perceived corruption.



The second team to present was the "CA" group, which showed  the gross domestic product (GDP), the Corruption Perceptions Index (CPI) and the foreign investments of the ten largest economies in the world in the interval of time from 2012 to 2017.

On the "X" axis they put the total Gross Domestic Product (Total GDP) data and in the "Y" axis, the data for Foreign Direct Investment.
They identified the level of Corruption Perception by color intervals: closer to the red are countries with high levels of perceived corruption and closer to the opposite green, less perceived corruption.

The foreign direct investment indicator is net investment inflows to acquire a long-term management interest (10\% or more of the
voting capital) in a company operating in an economy other than that of the investor.


We performed the evaluation of this Short Datathon using an adaptation of the Retrospective technique used in Scrum\footnote{https://www.scrum.org/. Last Access: Jan. 8th, 2019.}. We distributed two Post-it notes to each participant where they could write aspects that they "liked" on the event and aspects that they "did not like." The whiteboard of the room was divided into two parts ("I liked" and "I did not like") so that each participant could hang up the post-it with their comments.


We also applied surveys for collecting demographic information and including a section to obtain more information about the effectiveness of the
format as an aid in learning data analysis and visualization skills  (Likert five-point scale questions).

\section{Case Study II}

In the second case study we incorporated some changes from the experience with the first study. For example, we planned to give awards for the best presentations and we extended the time duration from 6 to 8 hours. In this second case study, the theme of the challenge was more specific.  We requested donations of nonperishable products to participate as a way to guarantee that those that signed up actually came to the event. 

Another aspect was the creation of a Facebook page for dissemination. In addition, we allowed participants to bring their own computer so they could use any software. Another unexpected novelty of this second study was the participation as a volunteer of a participant from the previous case study. This person received the invitation for the second event and, instead of participating in a team, preferred to help all teams with her experience in the tool.

Again we have used a laboratory with capacity for 25 people, with computers, chairs and projector. 

Some risks were also considered for the second case study. Unlike the previous event, organization and dissemination were independent. We chose a Saturday with the intention of attracting more participants that were not students.

We started Short Datathon at 9:15AM with the presence of 8 students and professionals from the areas of Computer Science, Economics and Business Administration. We contextualized the topic and explained the workings of Gapminder and then the participants were divided into teams. The teams carried out an initial exploration of the data and submitted their proposals.

After lunch break, 5 participants returned to continue with the work. At the end of the afternoon the teams presented their work. The "SEGA Data" team presented an Analysis of Informality in Brazil comparing informality data to other variables such as unemployment rate, corporate evasion and the broad consumer price index in an annual
interval from 2001 to 2014. In describing the problem, the team mentioned that informality in Brazil is one of the causes of tax evasion. It is common in the country that professionals provide services without formalization and the financial resources transacted may or may not go through the financial system.


The data sources used were from: Getulio Vargas Foundation (FGV), Institute of Applied Economic Research\footnote{http://www.ipeadata.gov.br/Default.aspx. Last access on Jan. 8th, 2019.}, Brazilian Institute of Geography and Statistics (IBGE)\footnote{https://www.ibge.gov.br/. Last Access: Jan. 8th, 2019.} and Brazilian Institute of Planning and Taxation (IBPT)\footnote{https://ibpt.com.br/. Last Access: Jan. 8th, 2019.}.

The second team, called "Olho de Lince", presented data on the Brazilian Federal Government spending between 2007 and 2018 on health, education, public safety and social security compared to the estimated amount of annual tax evasion. While spending on public services is high, they have concluded that budget revenues lost due to tax evasion could cover many the Brazilian Federal Government expenses.


The data source used was the National Treasury of the Ministry of Finance\footnote {http://www.tesouro.fazenda.gov.br/. Last access on Jan. 8th, 2019.}. To obtain an estimate of the amount withheld in local (Brazilian Real) currency, the team did a survey on the estimates made by the Internal Revenue Service and made one projection per year.

The instructor (first author) and the judge (third author) gave verbal feedback on the presentations but decided not to give any award, as both presentations were almost equally good.

We carried out the evaluation of the event through an online form sent by email to the participants and we also applied surveys for collecting demographic information and to obtain more information about the effectiveness of the format as a support in the development of data analysis and visualization skills.

\section{Results and Lessons Learned}

During the design and implementation of the Short Datathon, several key decisions were included to ensure the success of the events:
\begin{itemize}
	\item The first author and instructor of the activities was exposed to new topics related to Economics, Finance and Commerce which required previous exhaustive study;
	\item The first author made a major effort to include software tools that were available for free, had online learning resources and could be accessed independently outside the program so participants continued their research and analysis beyond the Short Datathon event.
	
\end{itemize}

Among the results that can be mentioned, we can list the participants considerations and the difficulties faced. The comments received
in the retrospective of the first case study were positive. Some of the comments received in the second case study were \textit {``I believe the experience was valid! I make a suggestion for the Environment theme for an upcoming event. Look for partnerships with NGOs, public bodies to be able to expand this event, opening space for greater participation of society as a whole"} and \textit {``The theme is very interesting and should continue to be addressed, because in a country like Brazil tax evasion is enormous and I believe that using computer science we can identify all fiscal deviations, money laundering, etc."}




\begin{table}[!h]
	\centering
	\caption{Table of Likert values of the cases study}
	\includegraphics[width=\linewidth]{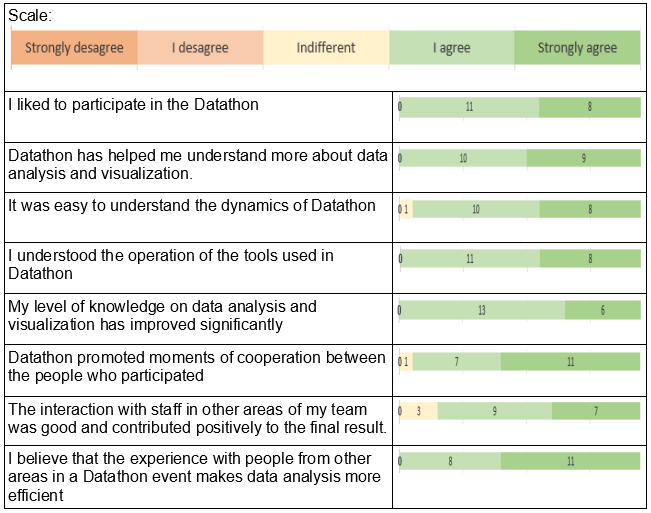}
	\label{tableLikertMod2}
\end{table}

In Table \ref{tableLikertMod2} we present the results of the part of the survey that included Likert scale statements. By analyzing the results obtained, we can mention that it was possible to identify similarities with a study carried out previously \cite {Nandi2016} because the results suggest that the constant collaboration between the participants of each team helps them achieve the final result. Short Datathons could assist as an informal platform for developing data analysis and visualization skills as in the experience mentioned by
\cite{Anslow2016}. 

Among the limitations of this study, we can mention that: 1) only two case studies were applied; 2) only specific topics of tax evasion were used for the challenge; 3) the number of participants was small.

\section{Conclusion}

In the present work we proposed a Short Datathon format for the development of data analysis and visualization skills. We prepared a schedule of activities for working with actual data on tax evasion.

We performed two case studies to evaluate the Short Datathon event format. The teams presented a total of 4 presentations of data analysis and visualization. In the questionnaire, the participants mentioned that Short Datathon helped them understand more about data analysis and visualization and that the format was understandable and promoted moments of cooperation between people. They also stated that interaction with people from other areas was good and contributed positively to the final result. The main limitation of this study was the number of participants.  Further studies are necessary to evolve the format of the event and better evaluate its effectiveness.


\section{Acknowledgments}

This work was supported by a PAEC-OEA-GCUB scholarship awarded to the first author by the Federal University of Technology - Paraná (UTFPR).

\bibliographystyle{IEEEtran} 
\bibliography{IEEEabrv,referencias}

\end{document}